# High quality MgB$_2$ thin films in-situ grown by dc magnetron sputtering

R. Vaglio, M.G. Maglione and R. Di Capua

I.N.F.M. and Dipartimento di Scienze Fisiche,
Università di Napoli *Federico II*, Napoli, ITALY

**Abstract**

Thin films of the recently discovered magnesium diboride (MgB$_2$) intermetallic superconducting compound have been grown using a magnetron sputtering deposition technique followed by in-situ annealing at 830°C. High quality films were obtained on both sapphire and MgO substrates. The best films showed maximum Tc = 35 K (onset), a transition width of 0.5 K, a residual resisitivity ratio up to 1.6, a low temperature critical current density Jc > 10$^6$ A/cm$^2$ and anisotropic critical field with $\gamma \cong 2.5$ close to the values obtained for single crystals.
The preparation technique can be easily scaled to produce large area in-situ films.

**Introduction**

The recently discovered intermetallic superconducting compound MgB$_2$ [1] presents very interesting aspects both related to its record Tc for non-oxide inorganic superconductors (40K), for the anisotropic superconducting properties and for the intriguing two-band electronic structure [2].
Thin films of MgB$_2$ are of special interest for microelectronic devices due to the relatively high Tc, the sufficiently long coherence lenght and the fairly high critical current density [3-5].
The availability of in-situ deposited, large area, high quality films with relatively simple techniques is of great importance for the development of such applications.
Unfortunately the high vapor pressure of Mg (respect to the Boron's one) even at low temperatures has made extremely difficult the realization of high quality thin films in a single step, without a Mg high pressure postanneal [6-8], extremely desirable for a large number of studies and applications.
As a second possibility, two-steps in-situ processes can be of interest in the same framework. Many groups have realized reasonable quality in-situ films by PLD deposition techniques and a subsequent high temperature annealing procedure [9-11], or by a single step procedure [12,13]. The results were encouraging though the film quality resulted to be inferior in respect to the best films obtained with ex-situ annealing treatments at much higher Mg partial pressures [6-13]. In our knowledge, the only MgB$_2$ thin films obtained without ex-situ annealing with Tc>30K are the ones reported in ref. [11] (in-situ annealing) and [13] (as grown). Furthermore, up to now, there are only few reports in the literature about MgB$_2$ thin films deposition by sputtering [14].
In the present paper we present a simple, innovative method to produce high quality films by a two-step, in-situ technique starting by a precursor film obtained by magnetron sputtering. Among other advantages, the proposed procedure can be easily scaled to obtain large area films for practical applications.

**Fabrication procedure and structural characterization**

MgB$_2$ thin films were grown at INFM - University of Naples by a planar magnetron sputtering technique in a UHV system operating at a base pressure in the low 10$^{-7}$ Pa range. Design, construction and testing of the balanced magnetron unit used are reported in ref. [15]. In brief, the system is equipped with 3 focused 2" magnetron sources. The substrates (sapphire or MgO) are placed "on axis" at variable distance from the target well outside the plasma region.

During the deposition of the precursor film the substrates were placed over a thin Mg disk, on the bottom of a cylindrical Nb box (h $\cong$ 4mm) placed on the surface of a molybdenum heater.

After the deposition, using tweezers mounted on a wobble stick, the Nb box is well closed by a properly designed cap. An indium wire gasket is inserted between box and cap. At this point the heater is switched on and ramped up to the desired temperature. The indium gasket melting guarantees a perfect sealing of the box. The box design and the internal Mg overpressure (due to the presence of the Mg disk) prevent any In diffusion in the film. The process is therefore conducted in saturated Mg-vapor as in ex-situ processes, giving high level film quality and reproducibility.

Best results were obtained codepositing $MgB_2$ and Mg for 10min on sapphire, resulting in a Mg rich Mg-B precursor film. The target voltages and current were 460 V ($MgB_2$ target), 400 V (Mg target) and 1 A respectively. The argon pressure during sputtering was $9 \times 10^{-1}$ Pa.

However this seems not to be at all critical and films with similar properties were obtained starting from precursors obtained sputtering only from the $MgB_2$ target. No attempts were made at the moment starting from B or B-rich Mg-B precursors.

Typical post-annealing processes were made at a substrate temperature $T_s = 830°C$ for 45 min (35 minutes were required to reach the equilibrium temperature).

After deposition the films could be removed from the system in the sealed Nb box for transfer in a UHV enviroment for surface studies (STM or others). The box can be also easily opened in situ with a properly designed device for further film processing (junction fabrication or other uses).

The process is highly reproducible and similar results were obtained on both sapphire and MgO substrates. The film thickness, measured with a standard stylus profilometer, were in the range (depending on deposition time and precursor composition) 0.8 - 1.0 μm.

XRD pattern (Cu Kα), reported in Fig. 1 for a film deposited on crystalline MgO(111), showed the $MgB_2$ phase with small amounts of MgO (probably due to the oxidation of Mg) and some unidentified spurious phases. By comparing the experimental spectra with the ones obtained on powders and polycrystalline samples, the relative height of the peaks seems to indicate a wide c-axis orientation of our films. From peaks positions, we can estimate the lattice parameters: a = b = 0.310 ± 0.002 nm, c = 0.353 ± 0.005 nm. X-rays on optimized films on sapphire are not reported because of the overlapping of the (00n) film reflections with the $Al_2O_3$ substrate reflections.

Film surface morphology was investigated by AFM. It shows a granular nature of the film surface. The roughness on a single grain (columnar nature, few microns wide) is about 50nm.

**Main normal state and superconducting properties**

In Fig. 2 the resistivity as a function of the temperature is reported for a typical film grown on a sapphire substrate in the standard conditions described in the previous section. The absolute value of resistivity was determined using a commercial linear four probe station, the main error being associated with effective thickness overestimation due to the strong granular nature of the samples surface. The estimated room temperature resistivity is about 200μΩcm, much higher in respect to single crystals. However, as discussed above, our value has only to be considered as an upper limit. Similar or higher values have been reported in literature for films [3,5], possibly due to similar effective thickness problems.

We do not have a clear explanation for the reduced onset critical temperature (35 K), possibly due to the somewhat low maximum annealing temperature that ca be reached inside our Nb box.

The transition width is $\Delta T_c = 0.5$ K (10%-90% criterion). Equivalent results were obtained for films grown on MgO substrates. In the temperature range explored, the resisitivity follows closely a Bloch-Gruneisen law. The continuous line in Fig. 2 represents the data best-fitting. From the best fit procedure, we obtained a Debye temperature $\Theta_D = 1100 ± 50$K. At low temperatures a clear $T^3$ dependence is observed as generally reported in the literature [5].

Our best films had a residual resistivity ratio (RRR) $\beta = \rho(300K)/\rho(40K) = 1.6$. A clear Tc-$\beta$ correlation has been observed in film of different quality, as reported in the inset of Fig. 2. The observed relation strongly resembles the behavior of many intermetallic classic superconducting compounds (Nb, A15) [5] as well as borocarbides [16].

$T_C$ and $J_C$ of our films were also evaluated inductively by measuring the third harmonic component voltage $V_3$ across a small sensor coil mounted very close to the film surface, as a function of temperature or driving current respectively [17]. The temperature dependence of the critical current performed by this method on our standard films is reported in Fig. 3. The extrapolated value for the zero field, zero temperature critical current is $Jc(0) = 1.6$ MA/cm$^2$ (Fig.1). However, best films exhibits a critical current density of 2MA/cm$^2$ at 11K. In the inset the Tc evaluation using a plot of $V_3$ vs T is also shown (as generally observed by this method the onset of the $V_3$ signal roughly corresponds to the zero resistance value temperature in resistive measurements).

In Fig. 4 the temperature dependence of the upper critical field both in the parallel and perpendicular direction with respect to the film plane are reported up to a maximum field of 8 Tesla. As typically reported [5] both curves show a slight upward curvature close to Tc. From the critical field slope we can estimate [18,19], at T = 0K, $\xi_\perp = 2.3 \pm 0.1$ nm and $\xi_\parallel = 5.7 \pm 0.2$ nm and an anisotropy ratio $\gamma = 2.5$ (slowly temperature-dependent). This value agrees fairly well with some current determinations on single crystals [20,21], indirectly proving the high quality and c-axis orientation of our films. A perfect fitting of the angular dependence of the critical field with the Ginzbug-Landau formula [22] was possible at all temperatures. Non optimized films presented a somewhat lower $\gamma$. Full critical field anisotropy measurements and analysis as well as critical current measurements vs. magnetic field will be reported elsewhere [23].

Finally, in Tab. I the main properties of our films are summarized. They can be well compared with the best films obtained up to now [14].

In conclusion we have presented a new method to produce high quality films by a two-step, in-situ technique starting by a precursor film obtained by magnetron sputtering. The fabrication process has to be still fully optimized but the overall film quality compares well to the best films obtained by two-step, ex-situ processes. The proposed preparation technique can be easily scaled to produce large area "in-situ" films for different applications.

**Acknowledgements**

The authors wish to thank L. Maritato and M. Salvato (INFM Salerno) for the critical field measurements and interpretation, C. Ferdeghini, M. Putti and V. Palmieri for discussions and suggestions, A. Cassinese and M. Salluzzo (INFM Napoli) for their continuous help during this work, G. Ausanio for the AFM measurements, F. Chiarella for the collaboration in the films fabrication. The technical support of A. Maggio and S. Marrazzo is also warmly acknowledged. This work has been partially supported by INFN-MABO.

**Figure Captions**

Fig. 1: a) X-ray diffraction (θ-2θ) spectrum for a $MgB_2$ film on MgO(111) substrate; b) comparison with a θ-2θ spectrum obtained on $MgB_2$ powders (from [6]).

Fig. 2: resistivity vs. temperature for a $MgB_2$ film grown on $Al_2O_3$ substrate. The continuos line represents the fit made using the generalized Bloch-Gruneisen formula. In the inset is shown the correlation between the Residual Resistivity Ratio (β) and Tc.

Fig. 4: critical current density vs. temperature for a $MgB_2$ film grown on $Al_2O_3$ substrate. The inset shows an inductive measurement of Tc.

Fig. 5: temperature dependence of the upper critical field both in the parallel and perpendicular direction with respect to the film plane for a $MgB_2$ film grown on $Al_2O_3$ substrate.

**Tab. I**

| | |
|---|---|
| Tc (K) | 35 |
| ΔTc (K) | 0.5 |
| β | 1.6 |
| ρ (μΩ cm) | 200 |
| Jc (MA/cm$^2$, 0 Field) | 2 (at 11K) |
| Hc$_2$ par (T) at T = 25.7K | 8.0 |
| Hc$_2$ perp (T) at T = 25.7K | 2.9 |

Tab. I: main properties of our best films.

**Figures**

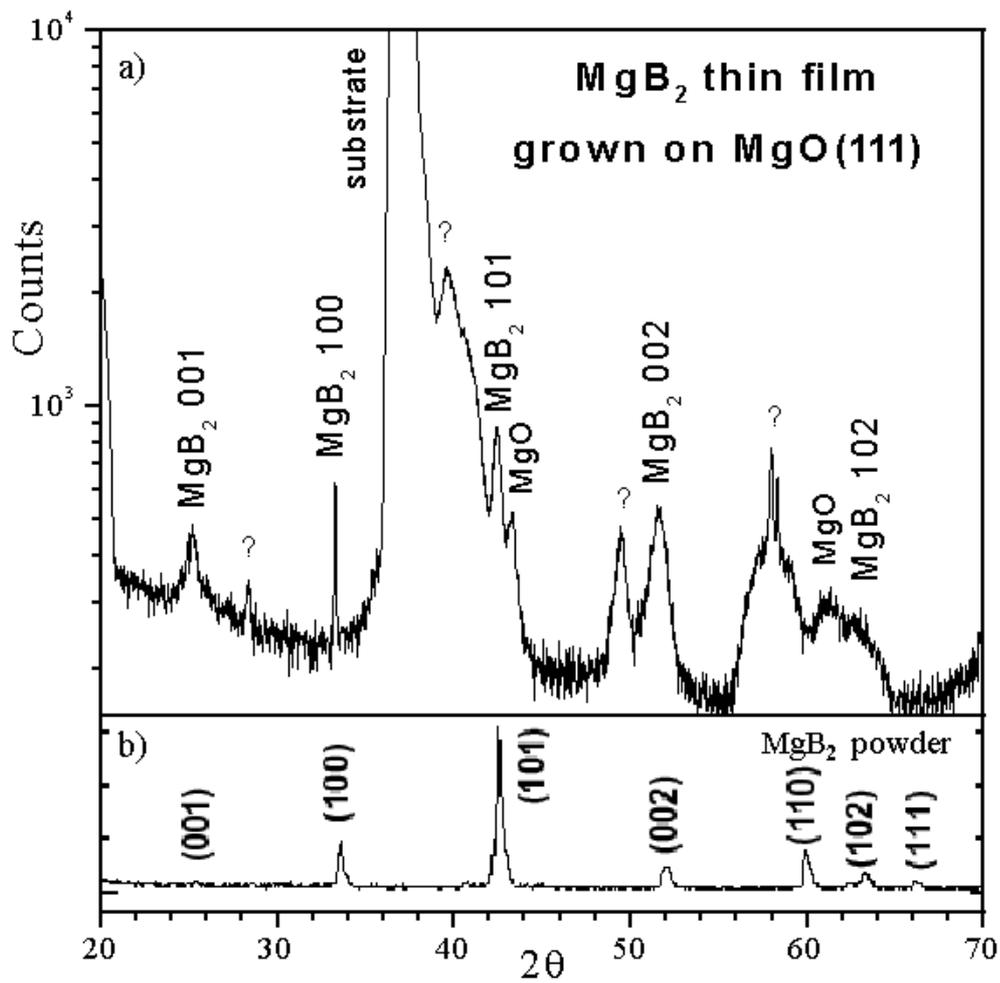

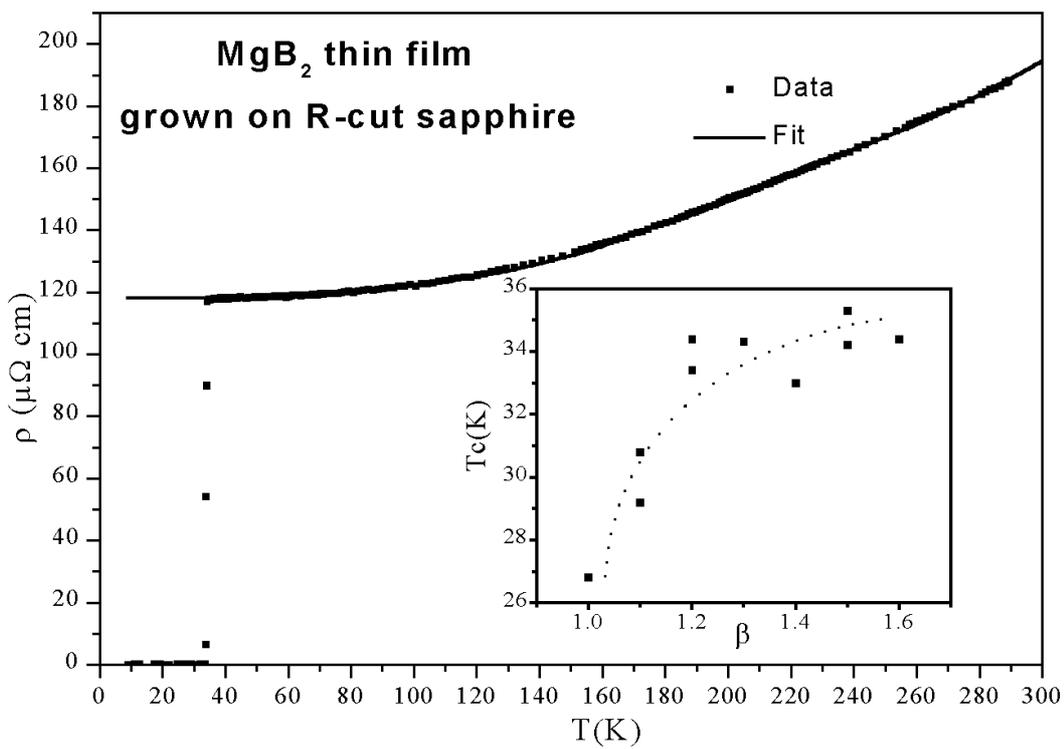

Fig. 1

Fig. 2

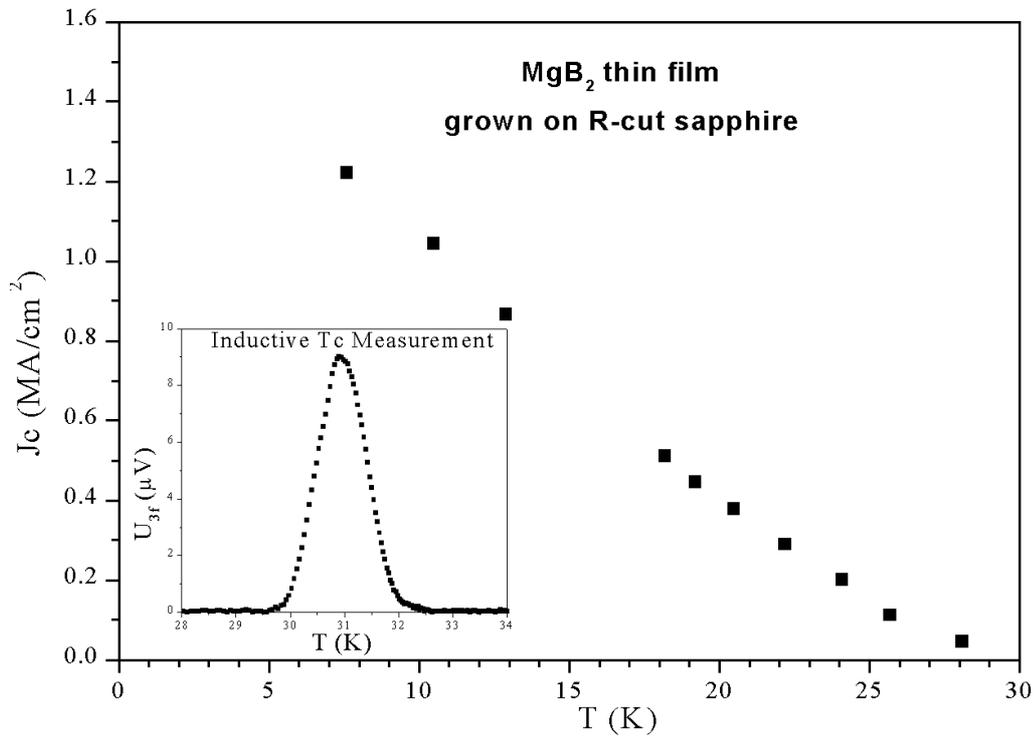

Fig. 3

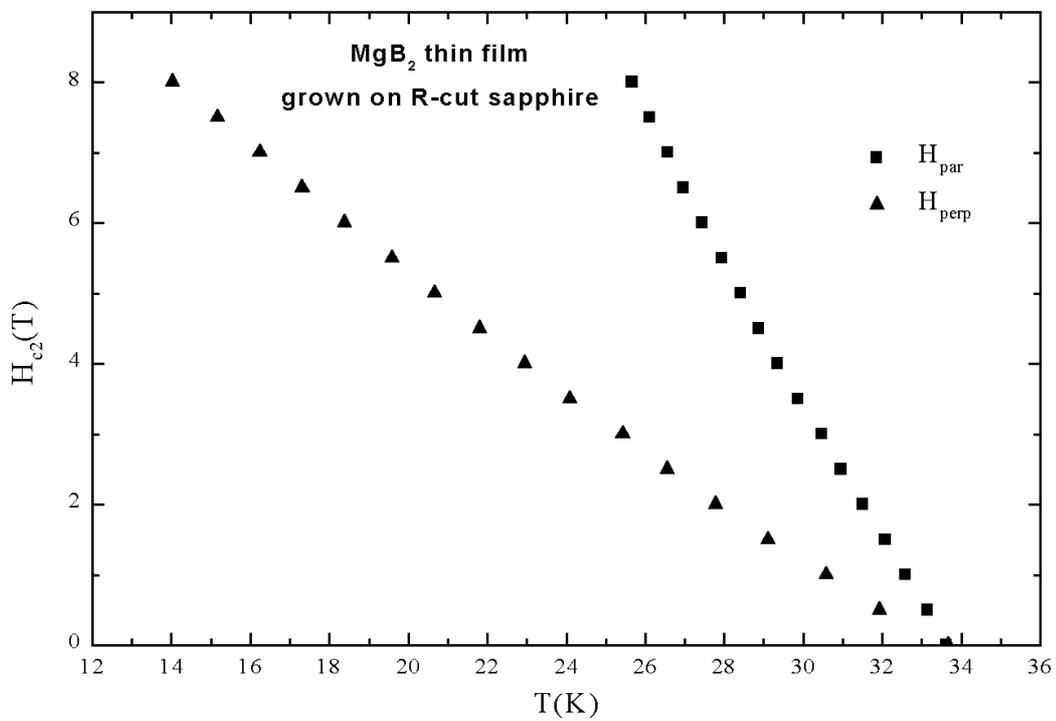

Fig. 4